\pdfoutput=1
\documentclass[11pt]{article}

\usepackage[final]{acl}

\usepackage{times}
\usepackage{latexsym}

\usepackage[T1]{fontenc}
\usepackage[utf8]{inputenc}

\usepackage{microtype}

\usepackage{inconsolata}

\usepackage{graphicx}

\usepackage{bm}
\usepackage{subfig}
\usepackage{amsmath}
\usepackage{booktabs}
\usepackage{multirow}
\usepackage{graphicx}
\usepackage{amssymb}
\usepackage{amsmath}
\usepackage{amsthm}
\usepackage{colortbl}
\usepackage{algorithm}
\usepackage{algorithmic}
\usepackage{subfig}

\usepackage{tcolorbox}
\tcbuselibrary{breakable}
\usepackage{wrapfig}
\usepackage{array}
\usepackage{longtable}
\title{Autoregressive Retriever: Improving Query Understanding from Item Feedback for Universal Multimodal Retrieval}

\author{
 \textbf{Jianfei Zhao\textsuperscript{1,2}},
 \textbf{Yifan Wang\textsuperscript{4}},
 \textbf{Feng Zhang\textsuperscript{1}},
 \textbf{Xin Sun\textsuperscript{1}},
 \textbf{Chong Feng\textsuperscript{1,5}},
 \textbf{Zhixing Tan\textsuperscript{3}}, \\
 \textbf{Yang Luo\textsuperscript{4}},
 \textbf{Boyuan Pan\textsuperscript{4}},
 \textbf{Xu Kai\textsuperscript{4}},
 \textbf{Yao Hu\textsuperscript{4}} \\
 \textsuperscript{1}School of Computer Science and Technology, Beijing Institute of Technology, \\
 \textsuperscript{2}Zhongguancun Academy,
 \textsuperscript{3}Zhongguancun Laboratory 
 \textsuperscript{4}Xiaohongshu \\
 \textsuperscript{5}Southeast Academy of Information Technology, Beijing Institute of Technology, \\
\small{
 {zhqingan}{@bit.edu.cn}
 }
}

\begin{document}
\maketitle
\begin{abstract}
Universal multimodal retrieval typically encodes a query once and ranks independently indexed items by embedding similarity. This design supports efficient search, but leaves the query representation unchanged even when retrieved items could help clarify the information need. We introduce the AutoRegressive Retriever (ARR), a multimodal retrieval model that learns both to select informative items and to use their content to refine subsequent retrieval. ARR alternates between retrieving an item and updating the query embedding, then uses the final embedding to rank the collection. Supervised fine-tuning teaches the encoder to use feedback through stepwise contrastive supervision.
Reinforcement learning treats feedback items as actions and optimizes their selection using the final reciprocal rank of a relevant item. A query-side adapter enables this optimization against a fixed item index.
ARR demonstrates strong retrieval performance on both in-domain and zero-shot benchmarks, outperforming the compared baselines on average.
Further analyses show that feedback improves retrieval at inference time and that training with feedback also improves the initial query embedding, before any item is observed.

\end{abstract}

\section{Introduction}

Retrieval systems increasingly need to search across text, images, and their combinations, supporting a wide range of applications, including retrieval-augmented generation~\citep{QueryAgent, Reasonrec}, multimedia platforms~\citep{UniNote}, and e-commerce search~\citep{OneRet, QueryAgent}. Universal multimodal retrieval (UMR) addresses this need with a single model that maps heterogeneous queries and items into a shared embedding space~\citep{M-BEIR, Lamra, MM-Embed}. In the standard embedding-based approach, queries and items are encoded independently and matched by similarity. This separation enables item embeddings to be computed in advance and reused across queries, making large collections searchable through a precomputed index.

% \begin{wrapfigure}{r}{0.5\linewidth}
\begin{figure}[!tbp]
    \centering
    \includegraphics[width=1\linewidth]{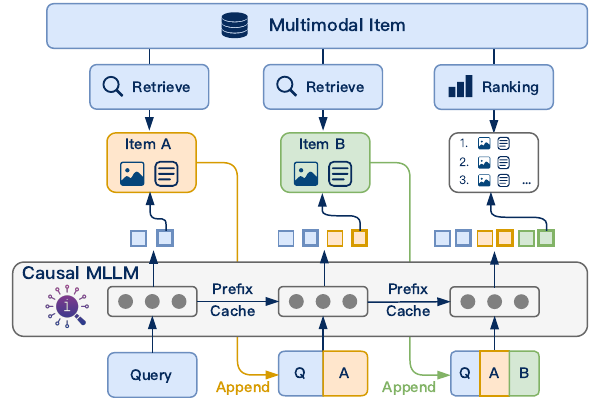}
    \caption{
    Autoregressive retrieval with ARR. Retrieved items provide context for successive query embeddings; the final embedding ranks the collection.
    }
    \label{fig:arr}
% \end{wrapfigure}
\end{figure}

However, this separation also limits a query's ability to leverage the collection being searched. A query may omit important details, express its intent only implicitly, or contain visual and textual cues of unequal importance. An independently computed query embedding must resolve such ambiguities before observing any candidate items. Although contrastive training \citep{moco, simclr} learns to align queries with relevant items across the training data, at inference time the query representation remains fixed and cannot adapt to the candidates encountered during retrieval. We refer to this lack of candidate-conditioned adaptation as the \emph{query--item gap}.

Recent works improve query interpretation through stronger multimodal large language models (MLLMs)~\citep{E5-V, Qwen3-VL-Emb}, richer representations~\citep{BToks, LaME}, and reasoning before embedding~\citep{TTE, TWN, TRACE}. These approaches extract more information from the input, but do not by themselves determine how a retriever should respond to newly encountered items. Retrieval feedback offers a complementary direction.
% Prior methods use retrieved documents to rewrite queries~\citep{GPRF, ADORE} or directly improve dense query representations~\citep{ANCE-PRF}.
Prior methods either directly refine dense query representations using retrieved documents~\citep{ANCE-PRF}, or rewrite queries and re-encode them~\citep{GPRF, ADORE}. Yet, the retrieved items are typically processed by an auxiliary refinement mechanism rather than by the retriever itself. We instead ask: \textbf{can the retriever learn to use the items it actually encounters to decide how to search better?}

We introduce the \textbf{A}uto\textbf{R}egressive \textbf{R}etriever (ARR), a multimodal retrieval model that alternates between retrieving an item and updating the query embedding (Figure~\ref{fig:arr}). We refer to this iterative process as \emph{autoregressive retrieval}. At each intermediate step, ARR selects an item by embedding similarity, appends its multimodal content to the query history, and encodes the expanded context for the next search. After a fixed number of steps, the final embedding ranks the collection. The process is autoregressive over retrieved items: each selection depends on the original query and earlier feedback. Item embeddings remain independently indexable, preserving the standard embedding-based search interface.

Learning this process requires distinguishing an item's relevance as an answer from its usefulness as feedback. An item that is not the target may still reveal information that helps retrieve the target later; conversely, a plausible match may distract the encoder. The model receives item content without a relevance label and must learn which evidence to use. We train ARR in two stages to address these challenges. Supervised fine-tuning (SFT) first teaches the encoder to interpret feedback using stepwise contrastive supervision on offline trajectories. Reinforcement learning (RL) then learns feedback selection from trajectories sampled by the current policy, using final retrieval quality as the reward. Its actions are retrieved items rather than generated text tokens. A query-side adapter learns this policy against a fixed item index.

Experiments on M-BEIR and seven zero-shot benchmarks show benefits from both feedback-conditioned inference and feedback-based training. ARR consistently achieves the highest average performance among all compared methods across both benchmarks. Training with feedback also improves first-step retrieval, indicating that its benefits extend to inference without feedback.

Our contributions are threefold:
\begin{itemize}
\item We introduce ARR, a multimodal retriever that performs autoregressive retrieval by conditioning query embeddings on a sequence of retrieved items while retaining independently indexed item embeddings.
\item We develop a two-stage training procedure that combines supervised feedback interpretation with item-selection policy optimization, linking intermediate retrieval actions to the final ranking outcome.
\item We evaluate ARR across in-domain and zero-shot tasks, and use ablation studies to distinguish the benefits of feedback at inference time from improvements learned during training.

\end{itemize}

\section{Related Work}

\paragraph{Universal multimodal retrieval.}
UMR brings diverse query--item modality configurations into a shared retrieval model~\citep{M-BEIR}. Contrastive models such as CLIP~\citep{CLIP} provide a foundation for cross-modal alignment, while MLLM-based encoders extend this approach to more flexible input combinations~\citep{E5-V, Qwen3-VL-Emb}. MM-Embed~\citep{MM-Embed} uses modality-aware hard-negative mining to improve multimodal retrieval. LamRA~\citep{Lamra} combines a retrieval model with a separate reranking component trained with pointwise and listwise objectives. BToks~\citep{BToks} and LaME~\citep{LaME} investigate bottleneck representations and latent computation for stronger embeddings. These directions improve representation learning or candidate ranking. ARR instead focuses on how retrieved content can inform subsequent query representations and how feedback items should be selected.

\paragraph{Retrieval feedback.}
Using retrieved results to refine a query is an established idea in information retrieval. ANCE-PRF~\citep{ANCE-PRF} learns a dense query encoder that consumes the original query and top-ranked documents while keeping the document index fixed. More recent approaches use language models to turn retrieval evidence into query reformulations: GPRF~\citep{GPRF} learns retrieval-oriented rewrites, and ADORE~\citep{ADORE} iteratively expands queries using assessments of retrieved content. ARR shares the motivation of adapting a query to the collection. Its focus is on learning a sequence of multimodal feedback actions within the embedding retriever: the encoder both consumes retrieved items and defines the policy that selects the next item, with selection trained for its effect on the final ranking.

\paragraph{Reasoning and reinforcement learning for retrieval.}
Reasoning-enhanced retrieval enriches inputs before constructing embeddings. TTE~\citep{TTE} uses generated context to improve multimodal representations, while TWN~\citep{TWN} and TRACE~\citep{TRACE} study when to invoke explicit reasoning. Retrieval-oriented query rewriting provides another way to expose information that is implicit in the input~\citep{zhu2024enhancing}. RL has also been used to align generation with retrieval outcomes: GRAPE~\citep{GRAPE} optimizes query rewriting using ranking rewards, and ELVA~\citep{ELVA} uses ranking and margin rewards to train representations produced after textual synthesis. ARR differs in its action space and feedback loop. It samples item identities directly from embedding similarities, observes the selected items, and optimizes these selections for downstream retrieval quality. Thus, its autoregressive sequence consists of retrieved items rather than a generated reasoning or rewriting trace.

\section{Preliminaries}
\paragraph{Task formulation.}
Let $\mathcal{D}$ denote an item corpus and $q$ a query. Both $q$ and an item $d\in\mathcal{D}$ may contain text, images, or a combination of the two. A query may also specify an instruction describing the desired relevance relation. UMR uses a shared retrieval model across these input configurations to rank items according to their relevance to $q$. We write a training example as $(q,d^+)$, where $d^+$ is an annotated relevant item.

\paragraph{MLLM-based encoder.}
An MLLM-based encoder maps queries and items into a shared embedding space. Text and images are converted into token representations and assembled using an input template. The MLLM processes the resulting sequence, and an embedding readout extracts a fixed-dimensional vector from its hidden states. After $\ell_2$ normalization, the embeddings and their similarity are
\begin{equation}
\begin{aligned}
        &\mathbf{z}_q=f_\theta(q),\qquad
    \mathbf{v}_d=f_\theta(d), \\
    &s_\theta(q,d)=\mathbf{z}_q^\top\mathbf{v}_d,
    \label{eq:similarity}
\end{aligned}
\end{equation}
where $f_\theta$ denotes the encoder and $\mathbf{z}_q,\mathbf{v}_d\in\mathbb{R}^{p}$. Their inner product measures cosine similarity in the shared embedding space.

\paragraph{Retrieval objective.}
The retriever is trained with a query-to-item contrastive objective. For each query $q$, let $\mathcal{B}_q$ contain its annotated positive $d^+$ and negative items. We use the InfoNCE loss~\citep{moco},
\begin{equation}\label{eq:infonce}
    \ell_{\mathrm{NCE}}\bigl(\mathbf{z},d^+,\mathcal{B}_q;\{\mathbf{v}_d\}\bigr)
    =-\log
    \frac{\exp(\mathbf{z}^{\top}\mathbf{v}_{d^+}/\tau)}
    {\sum_{d\in\mathcal{B}_q}\exp(\mathbf{z}^{\top}\mathbf{v}_d/\tau)},
\end{equation}
where $\tau>0$ is the temperature. Minimizing this loss increases the positive item's similarity relative to the negatives in $\mathcal{B}_q$.

\section{Method}
\label{sec:method}
ARR performs autoregressive retrieval by conditioning query representations on retrieved content while preserving independently encoded item embeddings. We first define this retrieval process (Section~\ref{ARR}), then describe SFT for interpreting feedback (Section~\ref{SFT}) and RL for selecting feedback that improves the final ranking (Section~\ref{RL}).

\subsection{Autoregressive Retrieval}\label{ARR}

ARR uses $N$ query states with $N-1$ feedback selections. It first encodes the original query, then repeatedly retrieves an item and appends its content and available metadata to the query context. The encoder processes this context to produce the next query embedding. The $N^\text{th}$ embedding determines the final retrieval results. No relevance label is included in the feedback.

% \begin{wrapfigure}{r}{0.32\linewidth}
\begin{tcolorbox}[
    title={Query Template}, width=\linewidth, breakable, colback=gray!4, colframe=gray!100, boxrule=1pt, arc=5pt, left=1pt, right=1pt, top=2pt,bottom=2pt, fontupper=\small
]
    \textbf{SYSTEM}: {\ttfamily <instruction>}
    % \medskip
    
    \textbf{USER}: {\ttfamily <query>}
    
    \textbf{ASSISTANT}: {\ttfamily <pad>}
    
    \textbf{USER}: {\ttfamily <item>}
    
    \textbf{ASSISTANT}: {\ttfamily <pad>}

    \ldots
    
\end{tcolorbox}
% \end{wrapfigure}

At step $t$, the model observes the original query and all previously selected feedback items. Let $a_t\in\mathcal{D}$ denote the item selected at step $t$. We define the state and query embedding as
\begin{equation}\label{eq:retrieval-state}
\begin{aligned}
    h_t=(q,a_1,\ldots,a_{t-1}),\quad\mathbf{z}_t=f_\theta\bigl(\mathcal{T}(h_t)\bigr),
    \\ t=1,\ldots,N,
\end{aligned}
\end{equation}
where $\mathcal{T}$ serializes the history using the MLLM's multi-turn conversation template, illustrated on the right. The \texttt{<pad>} token serves as the representation token.

A history mask prevents repeated selection of the same item. Let $\mathcal{C}(q)\subseteq\mathcal{D}$ denote the feedback candidate pool for query $q$, with at least $N-1$ items, and let $H_t=\{a_1,\ldots,a_{t-1}\}$ be the set of previously selected items. For $t<N$, embedding similarities define a policy over the remaining candidates:
\begin{equation}\label{eq:feedback-policy}
\begin{aligned}
&\pi_\theta(d\mid h_t;\mathcal{C}(q))
= \\[-0.5ex]
&\begin{cases}
\displaystyle
\frac{
\exp(\mathbf{z}_t^{\top}\mathbf{v}_d/\tau)
}{
\sum_{d'\in\mathcal{C}(q)\setminus H_t}
\exp(\mathbf{z}_t^{\top}\mathbf{v}_{d'}/\tau)
},
& d\in\mathcal{C}(q)\setminus H_t, \\[2ex]
0,
& \text{otherwise}.
\end{cases}
\end{aligned}
\end{equation}
% where $\tau_{\pi}>0$ is the policy temperature, distinct from the contrastive temperature $\tau$.
% During RL, we sample $a_t\sim\pi_\theta(\cdot\mid h_t;\mathcal{C}(q))$ and append its content to obtain $h_{t+1}$. At inference time, we select the highest-scoring unvisited item.
The probability of a sampled feedback trajectory $\xi=(a_1,\ldots,a_{N-1})$ factorizes as
\begin{equation}    \label{eq:trajectory}
    P_\theta(\xi\mid q;\mathcal{C}(q))
    =\prod_{t=1}^{N-1}
    \pi_\theta(a_t\mid h_t;\mathcal{C}(q)).
\end{equation}
This factorization makes retrieval autoregressive over item identities: earlier selections affect both the next query embedding and subsequent selection probabilities.
% After incorporating $N-1$ feedback items, ARR uses $\mathbf{z}_N$ to return top-rank items:
% $
%     \operatorname{Retrieve}_k(q,\xi)
%     =\operatorname{Top-}k_{d\in\mathcal{D}}
%     \mathbf{z}_N^{\top}\mathbf{v}_d.
% $
After incorporating $N-1$ feedback items, ARR uses the final query representation $\mathbf{z}_N$ to retrieve the top-$k$ items:
\[
\operatorname{Retrieve}_k(q,\xi)
=
\operatorname{Top\text{-}k}_{d\in\mathcal{D}}
\mathbf{z}_N^{\top}\mathbf{v}_d.
\]
The history mask is applied only to feedback selection; previously selected items remain eligible for retrieval at each step.

% The case $N=1$ reduces to single-pass embedding retrieval. For $N>1$, ARR adds query-encoding and search steps, with a growing feedback context, while retaining a precomputed item index. Its computational tradeoff is therefore additional query-side inference in exchange for candidate-conditioned adaptation.

\begin{figure*}[t]
    \centering
    \includegraphics[width=\linewidth]{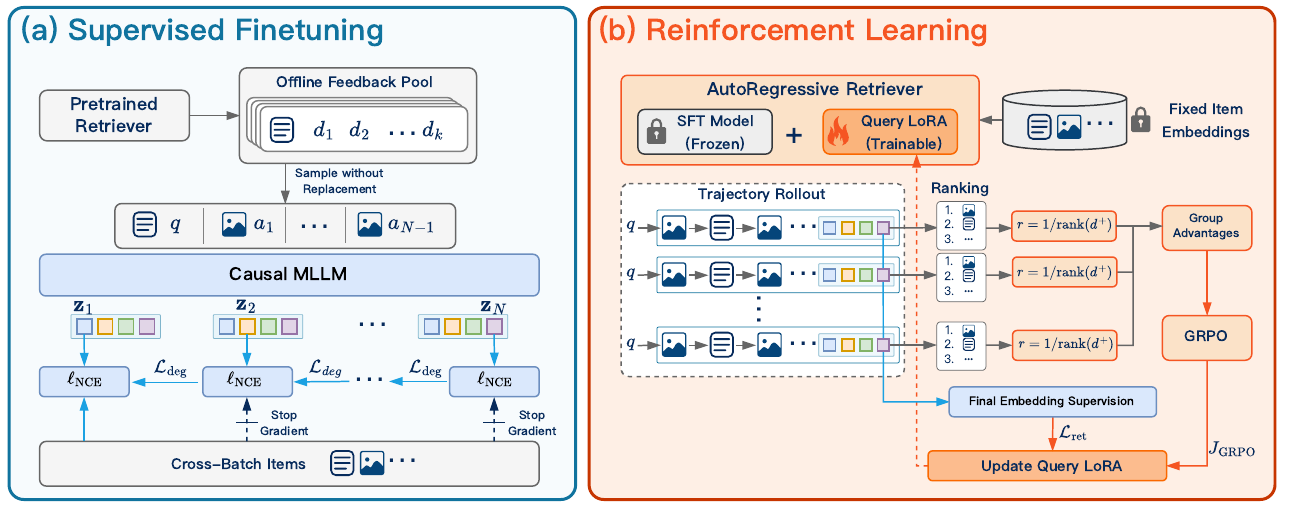}
    \caption{Overview of ARR training.
    \textbf{(a)} SFT uses offline feedback trajectories, stepwise contrastive supervision, and a degradation penalty. Only the initial-step loss backpropagates through standalone item embeddings.
    \textbf{(b)} RL freezes the SFT backbone and item index while training a query-side LoRA adapter. Terminal reciprocal-rank rewards supervise feedback selection through GRPO, complemented by contrastive supervision of the final embedding.
    }
    \label{fig:method}
\end{figure*}

\subsection{Supervised Fine-Tuning}\label{SFT}
SFT teaches the encoder to use retrieved content while preserving the original query intent (Figure~\ref{fig:method}a). To avoid repeatedly searching the corpus during this stage, we construct an offline feedback pool using the pretrained encoder with parameters $\theta_0$:
\begin{equation}
\mathcal{C}_{\mathrm{SFT}}(q)
=
\operatorname{Top-}K_{d\in\mathcal{D}}
f_{\theta_0}(q)^{\top}f_{\theta_0}(d). 
\end{equation}
Here, $K\geq N-1$. During training, we randomly sample an ordered sequence of $N-1$ items without replacement from this pool and construct the $N$ query states in Eq.~\ref{eq:retrieval-state}. We neither inject a missing positive into the feedback pool nor filter feedback by relevance labels. The contrastive set $\mathcal{B}_q$, which contains the annotated positive for supervision, is separate from this feedback pool.

We supervise every query embedding with the relevant item $d^+$. Let $\operatorname{sg}(\cdot)$ denote stop-gradient. We use the same contrastive set $\mathcal{B}_q$ across the states of a trajectory and define
\begin{equation}\label{eq:sft-step}
\begin{aligned}
        \ell_t=\ell_{\mathrm{NCE}}
    \bigl(\mathbf{z}_t,d^+,\mathcal{B}_q;
    \{\widetilde{\mathbf{v}}_d^{(t)}\}\bigr),
    \\
    \widetilde{\mathbf{v}}_d^{(t)}=
    \begin{cases}
        \mathbf{v}_d, & t=1,\\
        \operatorname{sg}(\mathbf{v}_d), & t>1.
    \end{cases}
\end{aligned}
\end{equation}
The first-step loss backpropagates through both query and item embeddings. Subsequent losses backpropagate through the query branch only, so feedback-conditioned states do not directly pull standalone item embeddings in different directions.
% Because the encoder parameters are shared, this does not freeze the item encoder throughout SFT.
% Gradients through feedback content inside the query context are also retained.

Retrieved content can clarify a query or distract the encoder. We therefore penalize increases in contrastive loss between consecutive states. For $N>1$, the degradation penalty is
\begin{equation}\label{eq:degradation}
\begin{aligned}
\mathcal{L}_{\mathrm{deg}}
&=
\frac{1}{N-1}\sum_{t=2}^{N}
\left[
\begin{aligned}
&\log(\ell_t+\delta) \\
&\quad-\log\bigl(\operatorname{sg}(\ell_{t-1})+\delta\bigr)
\end{aligned}
\right]_+ .
\end{aligned}
\end{equation}
where $[x]_+=\max(x,0)$ and $\delta>0$ is a numerical stability constant. The log difference measures a relative increase in loss. Taking the positive part at each step prevents an improvement elsewhere from canceling a harmful update. The stop-gradient prevents the reference loss from being directly increased to reduce the penalty. This encourages, but does not guarantee, improved rankings after feedback.

The overall objective is
\begin{equation}\label{eq:sft-objective}
    \mathcal{L}_{\mathrm{SFT}}
    =\mathbb{E}_{(q,d^+),\,\xi\sim p_{\mathrm{off}}(\cdot\mid q)}
    \left[\frac{1}{N}\sum_{t=1}^{N}\ell_t
    +\lambda_\mathrm{deg}\mathcal{L}_{\mathrm{deg}}\right],
\end{equation}
where $p_{\mathrm{off}}$ is the offline trajectory distribution and $\lambda_{\mathrm{deg}}\geq0$ weights the penalty. For $N=1$, we set $\mathcal{L}_{\mathrm{deg}}=0$. SFT supplies relevance supervision at every state without requiring expert labels for which feedback item to select.

\subsection{Reinforcement Learning for Retrieval}\label{RL}

SFT learns to interpret feedback, but does not explicitly optimize which items to select. Its trajectories come from an offline sampling procedure, and its contrastive losses supervise each state rather than assigning credit to actions for their downstream effects. We address these limitations by sampling trajectories from the current retrieval policy and optimizing feedback selection with a terminal ranking reward. Figure~\ref{fig:method}b illustrates this stage, which adapts group relative policy optimization (GRPO)~\citep{GRPO} to item-selection actions.

Let $\theta$ denote the parameters after SFT. We freeze this model and attach an additional query-side low-rank adaptation (LoRA)~\citep{LoRA} module with trainable parameters $\phi$. The query and item representations are
\begin{equation}\label{eq:rl-encoders}
    \mathbf{z}_t^{\phi}
    =f_{\theta,\phi}\bigl(\mathcal{T}(h_t)\bigr),
    \qquad
    \overline{\mathbf{v}}_d=f_{\theta}(d).
\end{equation}
The adapter is active only when encoding query contexts, including their feedback items. Standalone item embeddings are computed without it and remain fixed throughout RL. Consequently, optimization is confined to query interpretation and feedback selection within the SFT embedding space, without requiring the corpus index to be rebuilt after each update.

\paragraph{Terminal retrieval reward.}
The state is the complete history $h_t$, the action is a feedback item, and the transition appends that item to the history. This defines a finite-horizon Markov decision process with deterministic transitions. 
% To make training tractable, we use a query-specific pool $\mathcal{C}_{\mathrm{RL}}(q)$ containing the SFT model's top-100 candidates and the annotated positive. Feedback sampling, reward computation, and final-step contrastive supervision all use this pool.
Let $\mathcal{C}_{\mathrm{RL}}(q)$ denote the item pool for query $q$, and let $\{\overline{\mathbf{v}}_d\}_q$ denote the precomputed static item embeddings.
For a trajectory $\xi$, the terminal reward is the reciprocal rank of $d^+$ under the final query embedding:
\begin{equation}
    r_\phi(q,\xi)
    =1/\operatorname{rank}
    (d^+;\mathbf{z}_N^{\phi},\{\overline{\mathbf{v}}_d\}_q).
    \label{eq:reward}
\end{equation}
Rank is one-based and determined by descending similarity over the entire training pool, including any previously selected items.
% This local-pool reward is a training surrogate for ranking the full evaluation collection.

\paragraph{GRPO-style optimization.}
For each query, we sample $M\geq2$ trajectories $\{\xi^{(m)}\}_{m=1}^M$ from a rollout policy $\pi_{\phi_{\mathrm{old}}}$, using Eq.~\ref{eq:feedback-policy} over $\mathcal{C}_{\mathrm{RL}}(q)$ with fixed item vectors. We compute rewards $r_m=r_{\phi_{\mathrm{old}}}(q,\xi^{(m)})$ for each trajecory.

For an observed state--action pair, the probability ratio is
\begin{equation}
    \rho_{m,t}(\phi)=
    \frac{\pi_\phi(a_t^{(m)}\mid h_t^{(m)};\mathcal{C}_{\mathrm{RL}}(q))}
    {\pi_{\phi_{\mathrm{old}}}(a_t^{(m)}\mid h_t^{(m)};\mathcal{C}_{\mathrm{RL}}(q))}.
    \label{eq:policy-ratio}
\end{equation}
The objective for one query is
\begin{equation}
\begin{aligned}
    J_{\mathrm{GRPO}}(\phi;q)
    =&\frac{1}{M(N-1)}\sum_{m=1}^M\sum_{t=1}^{N-1}
    \min\Bigl[\rho_{m,t}(\phi)A_m, \\
    &\operatorname{clip}\bigl(\rho_{m,t}(\phi),
    1-\epsilon,1+\epsilon\bigr)A_m\Bigr].
\end{aligned}
    \label{eq:grpo-objective}
\end{equation}
Here, $\epsilon>0$ is the clipping parameter and $A_m=(r_m-\bar r)/\sigma_r$ is the group-relative advantage, where $\bar r$ and $\sigma_r$ are the mean and standard deviation of the $M$ rewards.
% When all rewards are equal, we take $A_m=0$, since the group provides no relative preference signal. 
% Advantages are held fixed during optimization.
Each action receives its trajectory's terminal advantage, encouraging selections associated with better final rankings within the group.

\paragraph{Final embedding supervision.}
The policy objective acts on the first $N-1$ states. Although the final embedding determines the reward, the detached ranking reward provides no gradient through that embedding. Shared parameters can still change its representation indirectly. We therefore add direct contrastive supervision at the final state, using the fixed item embeddings in $\mathcal{C}_{\mathrm{RL}}(q)$:
\begin{equation}
\label{eq:final-supervision}
\begin{aligned}
\mathcal{L}_{\mathrm{ret}}(\phi;q)
&=
\frac{1}{M}\sum_{m=1}^{M}
\mathbb{I}[r_m<R]
\\[-0.5ex]
&\quad \cdot
\ell_{\mathrm{NCE}}\Bigl(
\mathbf{z}_N^{\phi,(m)}, d^+,
\mathcal{C}_{\mathrm{RL}}(q);
\{\overline{\mathbf{v}}_d\}_q
\Bigr).
\end{aligned}
\end{equation}
% The gate is fixed during the update, as are the rollout rewards.
The combined objective is
\begin{equation}
\label{eq:rl-objective}
\begin{aligned}
\mathcal{L}_{\mathrm{RL}}(\phi)
&=
\mathbb{E}_{\substack{
q,\,\{\xi^{(m)}\}\\
\sim\pi_{\phi_{\mathrm{old}}}
}}
\left[
-J_{\mathrm{GRPO}}(\phi;q)
+\lambda\,\mathcal{L}_{\mathrm{ret}}(\phi;q)
\right],
\\[-0.5ex]
\lambda
&=
\lambda_{\mathrm{ret}}\,
\mathbb{I}[r<R].
\end{aligned}
\end{equation}
Here, $\mathbb{I}[\cdot]$ is the indicator function, $R$ is a reward threshold, and $\lambda_{\mathrm{ret}}\geq0$ weights the auxiliary loss. The gate is applied to each trajectory before averaging: only trajectories with rollout reward below $R$ receive the auxiliary loss. The gate remains fixed during the policy update. This concentrates direct supervision on trajectories with poorer final rankings.
We do not include a reference-policy KL penalty in the RL objective. Instead of constraining the policy to remain close to a separate reference model, we directly preserve retrieval quality through contrastive supervision on the final representation. This contrastive objective provides task-specific supervision for the final retrieval embedding, while the RL objective optimizes the model's feedback-selection behavior.

\section{Experiments}
\label{sec:experiments}
\subsection{Experimental Setup}
We present only the key experimental settings in this section, while Appendix~\ref{app:experimental-settings} provides the full details.

\paragraph{Benchmarks.}
We train and evaluate ARR on M-BEIR~\citep{M-BEIR}, which contains 16 dataset--task combinations spanning eight query--item modality configurations.

\paragraph{Implementation.}
We initialize ARR from the 2B and 8B Qwen3-VL-Embedding models~\citep{Qwen3-VL-Emb}. The visual encoder remains frozen throughout training. SFT fine-tunes the remaining backbone parameters, whereas RL freezes the SFT backbone and trains only a query-side LoRA adapter. SFT uses offline top-10 feedback pools. For RL, we sample 2K queries from each M-BEIR dataset--task combination, giving 32K queries in total.
To make training tractable, we use a query-specific pool $\mathcal{C}_{\mathrm{RL}}(q)=\mathcal{C}_{100}(q)\cup\{d^+\}$, where $\mathcal{C}_{100}(q)$ contains the SFT model's top-100 candidates. Feedback sampling, reward computation, and final-step contrastive supervision all use this pool.
Unless otherwise stated, we use $N=4$ for training and inference and select feedback greedily at inference time. ARR-SFT and ARR-RL denote the models after the two training stages.

\paragraph{Baselines.}
We compare against MM-Embed~\citep{MM-Embed} and LamRA-Ret~\citep{Lamra}, as well as TRACE~\citep{TRACE}, which incorporates adaptive reasoning, and ELVA~\citep{ELVA}, which uses ranking-driven RL. These comparisons cover contrastive embedding learning, reasoning-enhanced retrieval, and retrieval-oriented policy optimization.

\subsection{In-Domain Evaluation}

\begin{table*}[t]
\centering
\caption{M-BEIR results in the local-pool setting (\%).
Superscripts t and i denote text and image inputs, respectively.
ARR rows report the final state ($N=4$).
% Bold marks the best result in each column within each model-size group.
}
\label{tab:mbeir}
\setlength{\tabcolsep}{2pt}
\renewcommand{\arraystretch}{1.15}
% \fontsize{7}{8}\selectfont

\resizebox{\linewidth}{!}{%
\begin{tabular}{@{}l*{17}{c}@{}}
\toprule
& \multicolumn{3}{c}{$q^\text{t} \to d^\text{i}$}
& $q^\text{t} \to d^\text{t}$
& \multicolumn{2}{c}{$q^\text{t} \to d^\text{i,t}$}
& \multicolumn{3}{c}{$q^\text{i} \to d^\text{t}$}
& $q^\text{i} \to d^\text{i}$
& \multicolumn{2}{c}{$q^\text{i,t} \to d^\text{t}$}
& \multicolumn{2}{c}{$q^\text{i,t} \to d^\text{i}$}
& \multicolumn{2}{c}{$q^\text{i,t} \to d^\text{i,t}$}
& \\
\cmidrule(lr){2-4} \cmidrule(lr){5-5} \cmidrule(lr){6-7}
\cmidrule(lr){8-10} \cmidrule(lr){11-11} \cmidrule(lr){12-13}
\cmidrule(lr){14-15} \cmidrule(lr){16-17}
\textbf{Method} & \textbf{VN} & \textbf{CO} & \textbf{F200} & \textbf{WQ} & \textbf{ES} & \textbf{WQ} & \textbf{VN} & \textbf{CO} & \textbf{F200} & \textbf{NS} & \textbf{ON} & \textbf{InS} & \textbf{FIQ} & \textbf{CR} & \textbf{ON} & \textbf{InS} & \textbf{Avg.} \\
& R@5 & R@5 & R@10 & R@5 & R@5 & R@5 & R@5 & R@5 & R@10 & R@5 & R@5 & R@5 & R@10 & R@5 & R@5 & R@5 & \\
\midrule
\multicolumn{18}{@{}l}{\textit{2B models}} \\
LamRA-Ret~\citep{Lamra} & 30.8 & 78.8 & 23.1 & 82.5 & 54.3 & 77.8 & 31.2 & 88.5 & 27.1 & 28.7 & 51.1 & 44.2 & 28.9 & 47.7 & 72.3 & 60.8 & 51.6 \\
ELVA~\citep{ELVA} & \textbf{35.6} & 80.3 & 25.0 & 88.0 & 56.1 & 80.5 & \textbf{33.4} & 90.2 & 25.9 & 29.3 & 52.0 & \textbf{47.4} & 30.9 & 50.0 & 72.8 & 61.3 & 53.8 \\
\rowcolor{gray!20}
ARR-SFT & 32.6 & \textbf{84.4} & 27.2 & 93.3 & 60.4 & 84.1 & 32.1 & \textbf{94.3} & 27.6 & 32.2 & 55.2 & 42.9 & 33.1 & \textbf{59.7} & 74.2 & 62.8 & 56.0 \\
\rowcolor{gray!20}
ARR-RL & 32.0 & 84.0 & \textbf{28.2} & \textbf{94.0} & \textbf{63.9} & \textbf{85.1} & 31.8 & 94.2 & \textbf{28.7} & \textbf{34.1} & \textbf{57.7} & 42.4 & \textbf{33.3} & 59.3 & \textbf{75.3} & \textbf{63.9} & \textbf{56.7} \\
\midrule
\multicolumn{18}{@{}l}{\textit{8B/7B models}} \\
MM-Embed~\citep{MM-Embed} & 41.0 & 71.3 & 17.1 & \textbf{95.9} & 68.8 & 85.0 & 41.3 & 90.1 & 18.4 & 32.4 & 42.1 & 42.3 & 25.7 & 50.0 & 64.1 & 57.7 & 52.7 \\
LamRA-Ret~\citep{Lamra} & 41.6 & 81.5 & 28.7 & 86.0 & 62.6 & 81.2 & 39.6 & 90.6 & 30.4 & 32.1 & 54.1 & 52.1 & 33.2 & 53.1 & 76.2 & 63.3 & 56.6 \\
TRACE~\citep{TRACE} & 42.1 & 82.3 & \textbf{30.5} & 87.8 & 64.1 & 82.5 & 41.2 & 91.3 & \textbf{33.2} & 33.6 & 57.4 & 55.8 & \textbf{36.4} & 57.3 & \textbf{78.5} & 67.1 & 58.8 \\
ELVA~\citep{ELVA} & \textbf{43.5} & 83.0 & 29.2 & 91.0 & 63.5 & 83.1 & \textbf{41.7} & 92.2 & 32.1 & 32.8 & 56.0 & 55.5 & 34.6 & 55.4 & 77.5 & 67.1 & 58.7 \\
\rowcolor{gray!20}
ARR-SFT & 40.6 & \textbf{85.8} & 27.8 & \textbf{95.9} & 66.8 & 86.0 & 40.1 & \textbf{94.7} & 28.1 & 33.6 & 58.7 & \textbf{56.1} & 35.0 & \textbf{64.4} & 75.3 & 67.8 & 59.8 \\
\rowcolor{gray!20}
ARR-RL & 41.3 & 85.6 & 30.0 & 95.7 & \textbf{73.2} & \textbf{86.2} & 41.0 & \textbf{94.7} & 30.2 & \textbf{34.0} & \textbf{63.5} & 55.8 & 35.1 & 64.0 & 78.1 & \textbf{68.2} & \textbf{61.0} \\
\bottomrule
\end{tabular}%
}
\end{table*}

Table~\ref{tab:mbeir} reports results on the M-BEIR test set. ARR-RL achieves the highest average score among the compared methods at both scales: 56.7 for 2B and 61.0 for 8B. RL improves the reported averages over ARR-SFT by 0.7 and 1.2 points, respectively. At 8B, ARR-RL exceeds the strongest baseline average, TRACE's 58.8, by 2.2 points.

The gains vary across tasks. For the 8B model, RL increases EDIS (ES) text-to-multimodal recall from 66.8 to 73.2 and OVEN (ON) multimodal-to-text recall from 58.7 to 63.5. These tasks require connecting the query to an item with a different information structure, making them promising settings for feedback-conditioned adaptation.
% However, improvements are not universal: ARR-RL remains below the strongest baselines on VisualNews and Fashion200K, and RL slightly reduces recall on some tasks, including CIRR. The results therefore support an aggregate benefit from policy training rather than uniform improvement across datasets.
Overall, RL improves the SFT model on most tasks, demonstrating the effectiveness of policy optimization for ARR. Some task configurations exhibit slight performance degradation, including VisualNews (VN) at 2B and CIRR (CR) at both scales. We hypothesize that these drops may stem from the sensitivity of RL training to the data: since RL is performed intensively on a relatively small subset of training data, variations in data quality and distribution across tasks may introduce task-specific biases. Nevertheless, the gains on most tasks indicate that RL provides an overall improvement over the SFT model.

\subsection{Analysis of the Iterative Retrieval Process}

Table~\ref{tab:iteration} separates the effects of training with feedback from using feedback at inference time. Subscripts indicate the number of training states; columns report retrieval from each inference state. All RL-stage variants start from ARR-SFT$_{N=4}$. The $N=1$ RL-stage control receives additional contrastive training on the RL data but uses neither feedback nor GRPO during that training.

\paragraph{Feedback use must be learned.}
ARR-SFT$_{N=1}$ scores 55.23 at the initial state but falls to 46.90 when extended to four inference states, showing that simply appending retrieved content can harm a model not trained to use it. In contrast, ARR-SFT$_{N=4}$ improves from 55.63 to 56.01. Most of this gain occurs after the first feedback item, at Iter-2 (55.95), with smaller changes thereafter.

% \begin{wraptable}{r}{0.5\textwidth}
\begin{table}[!htbp]

\centering
\caption{
Effect of training and inference steps for ARR-2B on M-BEIR (Avg.\%).
% Subscripts give the training horizon; columns give inference states.
% All RL-stage variants start from ARR-SFT$_{N=4}$.
% The $N=1$ control uses the RL data with contrastive training and no feedback or GRPO during training.
% Bold marks column maxima within each block.
}
\label{tab:iteration}
\footnotesize
\setlength{\tabcolsep}{2pt}
\renewcommand{\arraystretch}{1.12}

\begin{tabular*}{0.9\linewidth}{@{\extracolsep{\fill}}l*{4}{c}@{}}
\toprule
\textbf{Method} & \textbf{Iter-1} & \textbf{Iter-2} & \textbf{Iter-3} & \textbf{Iter-4} \\
\midrule
ARR-SFT$_{N=1}$ & 55.23 & 49.85 & 48.37 & 46.90 \\
ARR-SFT$_{N=4}$ & \textbf{55.63} & \textbf{55.95} & \textbf{56.01} & \textbf{56.01} \\
\midrule
ARR-RL$_{N=1}$ & 55.44 & 55.57 & 55.54 & 55.47 \\
ARR-RL$_{N=2}$ & \textbf{56.40} & 56.59 & 56.58 & 56.64 \\
ARR-RL$_{N=3}$ & \textbf{56.40} & 56.57 & 56.62 & 56.63 \\
ARR-RL$_{N=4}$ & 56.36 & \textbf{56.68} & \textbf{56.66} & \textbf{56.74} \\
\bottomrule
\end{tabular*}
% \end{wraptable}
\end{table}

\begin{table*}[t]
\centering
\caption{Zero-shot retrieval results for 8B models (\%).
Iter-1 uses the initial query embedding without feedback; unqualified ARR rows use the final state ($N=4$).
An asterisk marks benchmarks with images sourced from COCO or FashionIQ.
% Bold marks the best result in each column.
}
\label{tab:zero-shot}
\setlength{\tabcolsep}{3pt}
\renewcommand{\arraystretch}{1.15}
\footnotesize

\resizebox{\linewidth}{!}{%
\begin{tabular}{@{}l*{11}{c}@{}}
\toprule
& \multicolumn{3}{c}{$q^\text{t} \to d^\text{i}$}
& \multicolumn{3}{c}{$q^\text{i} \to d^\text{t}$}
& \multicolumn{2}{c}{$q^\text{i,t} \to d^\text{i}$}
& $q^\text{dialog} \to d^\text{i}$
& $q^{\text{i} \oplus \text{t}} \to d^\text{i}$
& \\
\cmidrule(lr){2-4} \cmidrule(lr){5-7} \cmidrule(lr){8-9}
\cmidrule(lr){10-10} \cmidrule(lr){11-11}
\textbf{Method} & \textbf{Share4V} & \textbf{Urban*} & \textbf{Flickr} & \textbf{Share4V} & \textbf{Urban*} & \textbf{Flickr} & \textbf{CIRCO*} & \textbf{GeneCIS*} & \textbf{VisD*} & \textbf{MT-FIQ*} & \textbf{Avg.} \\
& R@1 & R@1 & R@1 & R@1 & R@1 & R@1 & MAP@5 & R@1 & R@1 & R@5 & \\
\midrule
LamRA-Ret~\citep{Lamra} & 93.3 & 95.1 & 82.8 & 88.1 & 94.3 & 92.7 & 33.2 & 18.9 & 62.8 & 60.9 & 72.21 \\
TRACE~\citep{TRACE} & 94.9 & 94.8 & 84.5 & 89.1 & 94.1 & 94.5 & 34.8 & 20.5 & 65.4 & 63.2 & 73.58 \\
ELVA~\citep{ELVA} & 96.6 & 96.1 & 84.4 & 92.0 & 95.5 & 95.2 & 34.5 & 20.2 & 65.3 & 61.2 & 74.10 \\
\midrule
ARR-SFT (Iter-1) & \textbf{98.8} & 98.9 & 87.3 & \textbf{99.0} & \textbf{99.0} & 97.7 & 39.3 & 20.9 & 75.5 & 63.1 & 77.95 \\
ARR-SFT & \textbf{98.8} & \textbf{99.1} & 87.6 & \textbf{99.0} & \textbf{99.0} & 97.8 & 41.6 & 21.0 & 76.3 & 63.8 & 78.39 \\
ARR-RL (Iter-1) & 98.4 & 99.0 & 87.5 & \textbf{99.0} & \textbf{99.0} & 98.0 & 39.9 & 21.0 & 76.6 & 65.9 & 78.42 \\
ARR-RL & 98.4 & 99.0 & \textbf{88.0} & \textbf{99.0} & \textbf{99.0} & \textbf{98.3} & \textbf{43.4} & \textbf{21.1} & \textbf{77.9} & \textbf{66.0} & \textbf{79.01} \\
\bottomrule
\end{tabular}%
}
\end{table*}

\paragraph{Feedback training also improves initial retrieval.}
Before observing any feedback, ARR-SFT$_{N=4}$ outperforms ARR-SFT$_{N=1}$ by 0.40 points. Thus, training on feedback-conditioned states benefits the shared encoder even when inference stops after the first step. One possible explanation is that supervising several contexts encourages features that also help represent the original query; these results establish the performance effect, rather than a particular feature-learning mechanism.

\paragraph{Policy training provides further gains.}
Policy training consistently improves upon the SFT initialization. ARR-RL$_{N=4}$ achieves scores of 56.36 at Iter-1 and 56.74 at Iter-4, compared with 55.63 and 56.01, respectively, for its SFT initialization. Notably, the improvement at Iter-1 indicates that the benefits of RL are not limited to exploiting feedback from previous retrieval steps.
We attribute this gain to the richer supervision provided by the RL objective. While the SFT objective treats each item as either positive or negative, the RL objective assigns credit to a selected item according to its contribution to the eventual retrieval outcome. It therefore provides a softer and more informative learning signal, allowing the model to favor useful intermediate items even when they are not annotated positives and to better distinguish hard negatives based on their utility for subsequent retrieval.
The performance gap between ARR-RL$_{N=1}$ and ARR-RL$_{N=2}$ further supports this interpretation. ARR-RL$_{N=1}$, which is optimized only with a contrastive objective, achieves 55.44 at Iter-1, whereas ARR-RL$_{N=2}$, trained with the RL objective, reaches 56.40 at the same step. These results suggest that policy-level credit assignment provides complementary supervision beyond conventional contrastive learning.

\paragraph{Additional steps yield diminishing returns.}
Training with two, three, or four states gives similar initial scores. Four-state training produces the best final score, but its advantage over two-state training at Iter-4 is only 0.10 points.
Together with the marginal improvements beyond Iter-2, these results suggest a favorable trade-off between test-time scaling and retrieval performance in ARR, with Iter-2 offering a good balance between computational cost and performance gains.

\subsection{Zero-Shot Evaluation}
Table~\ref{tab:zero-shot} evaluates the 8B models on seven benchmarks without task-specific fine-tuning.
% Some benchmarks reuse images from COCO or FashionIQ, as marked in the table; zero-shot therefore refers to transfer to these evaluation tasks and query formats, not a guarantee of disjoint image content.
ARR-RL obtains the highest reported average, 79.01, compared with 74.10 for the strongest baseline average. Its scores of 77.9 on Visual Dialog and 66.0 on Multi-round FashionIQ also indicate transfer to dialogue and interactive retrieval.

The models from both training stages benefit from feedback at inference time: ARR-SFT improves from 77.95 at Iter-1 to 78.39 at the final step, and ARR-RL improves from 78.42 to 79.01. The gains are most visible on CIRCO and Visual Dialog. RL also raises the average at both the initial and final states relative to SFT, although some individual scores decrease slightly. These results extend the two observed benefits of feedback---better inference with retrieved context and better initial representations after training---beyond the M-BEIR evaluation tasks.

\subsection{Ablation Study}

Table~\ref{tab:ablation} examines ARR's training objectives using the 2B model.
% All scores are averages over the M-BEIR local-pool tasks.

\paragraph{SFT design choices.}
% Allowing gradients through standalone item embeddings at feedback-conditioned steps lowers the final score from 56.01 to 55.71.
Retrieval feedback is intended to help the model better understand the query. Accordingly, contrastive supervision at feedback-conditioned steps should provide a learning signal to the query representation, and we therefore stop gradients through the standalone item embeddings. When item gradients are enabled, the final score drops from 56.01 to 55.71, suggesting that the model may exploit undesirable shortcuts to minimize the training loss rather than learn meaningful item representations.
Removing the degradation penalty leaves initial performance nearly unchanged (55.69 vs. 55.63), but eliminates the gain from feedback: the final score is 55.68 instead of 56.01. The penalty therefore improves the usefulness of later states rather than simply increasing first-step accuracy.

% \begin{wraptable}{r}{0.5\linewidth}
\begin{table}[!htbp]
\centering
\caption{
Training ablations for ARR-2B on M-BEIR (Avg.\%).
``w/o reward gate'' applies final-step supervision to all trajectories.
% Bold marks column maxima within each block.
}
\label{tab:ablation}
\footnotesize
\setlength{\tabcolsep}{2pt}
\renewcommand{\arraystretch}{1.12}

\begin{tabular*}{0.9\linewidth}{@{\extracolsep{\fill}}l*{4}{c}@{}}
\toprule
\textbf{Setting} & \textbf{Iter-1} & \textbf{Iter-2} & \textbf{Iter-3} & \textbf{Iter-4} \\
\midrule
ARR-SFT & 55.63 & \textbf{55.95} & \textbf{56.01} & \textbf{56.01} \\
\quad\emph{w/o} $\operatorname{sg}(\mathbf{v}_d)$ & 55.31 & 55.62 & 55.72 & 55.71 \\
\quad\emph{w/o} $\mathcal{L}_{\mathrm{deg}}$ & \textbf{55.69} & 55.61 & 55.66 & 55.68 \\
\midrule
ARR-RL & 56.36 & \textbf{56.68} & \textbf{56.66} & \textbf{56.74} \\
\quad\emph{w/o} $\mathcal{L}_{\mathrm{ret}}$ & 56.20 & 56.40 & 56.39 & 56.43 \\
\quad\emph{w/o} reward gate & \textbf{56.39} & 56.55 & 56.52 & 56.39 \\
\quad\emph{w/} KL & 56.20 & 56.46 & 56.50 & 56.49 \\
\quad\emph{w/o} $J_{\mathrm{GRPO}}$ & 55.74 & 55.65 & 55.47 & 55.29 \\
\bottomrule
\end{tabular*}
\end{table}

\paragraph{Policy and final embedding supervision.}
Removing the final embedding contrastive loss lowers the final score from 56.74 to 56.43, which remains above the SFT score. Conversely, removing GRPO yields 55.29 at the final step, below the scores of both the full RL model and its SFT initialization.
Because the final embedding is conditioned on the retrieval trajectory, supervising only the final embedding can disrupt the model's retrieval policy, which in turn degrades the quality of the final representation. This result further suggests that the gains of ARR-RL stem from learning an effective retrieval policy, rather than from exploiting superficial item-specific patterns.
The two objectives are therefore complementary in this setting: policy training improves feedback selection, and direct terminal supervision improves the representation used for ranking.

\paragraph{Reward gating and KL regularization.}
Applying the auxiliary contrastive loss to every trajectory, instead of gating it by reward, gives a final score of 56.39. This supports concentrating the auxiliary loss on trajectories with lower rewards. The variant with a KL penalty scores 56.49, compared with 56.74 for the default model. Thus, the default objective performs better in this comparison while avoiding a separate reference-model evaluation. Appendix~\ref{sec:hyperparameters} further examines the weights $\lambda_{\mathrm{deg}}$ and $\lambda_{\mathrm{ret}}$.

\section{Conclusion}
We presented ARR, a multimodal retriever that learns to select and use retrieved items as feedback through an autoregressive retrieval process. ARR preserves independently indexed item embeddings while conditioning the query on a sequence of retrieved observations. SFT teaches feedback interpretation, and RL optimizes item selection for the final ranking outcome. Experiments show improvements in average in-domain and zero-shot retrieval performance. Further analyses highlight two complementary benefits of item feedback: using feedback during inference improves subsequent retrieval, while learning from feedback during training improves initial retrieval even before any feedback is observed.

\bibliography{custom}
\clearpage
\appendix
\section{Detailed Experimental Settings}
\label{app:experimental-settings}

\subsection{Benchmarks}
\label{app:data}

\paragraph{In-domain evaluation.}
We use the M-BEIR local-pool setting~\citep{M-BEIR}, in which candidates from the task-specific collection are ranked for each query. M-BEIR comprises 16 dataset--task combinations from 10 datasets, covering eight query--item modality configurations and four domains: news, fashion, Wikipedia, and miscellaneous content. Table~\ref{tab:app-mbeir-data} lists rounded benchmark split sizes and candidate-pool sizes. We report Recall@10 for Fashion200K and FashionIQ and Recall@5 for all other tasks.

\paragraph{Zero-shot evaluation.}
Table~\ref{tab:app-zero-data} lists query counts, candidate-pool sizes, and metrics for the seven benchmarks in Table~\ref{tab:zero-shot}. ShareGPT4V and Urban-1K evaluate retrieval between images and long captions, and Flickr30K evaluates image--caption retrieval. CIRCO and GeneCIS condition retrieval on an image and text. Visual Dialog evaluates dialogue-to-image retrieval, while Multi-round FashionIQ uses multi-turn image--text interactions. We evaluate these tasks without task-specific fine-tuning. Several benchmarks reuse images from COCO or FashionIQ, so this protocol does not imply that all visual content is unseen.

\subsection{Implementation and Training}

\paragraph{Model and feedback pools.}
We use the 2B and 8B Qwen3-VL-Embedding backbones~\citep{Qwen3-VL-Emb}. The visual encoder remains frozen in both stages. SFT fine-tunes the remaining backbone parameters; RL freezes the SFT backbone and trains an additional query-side LoRA adapter. The adapter is active when encoding the query and its feedback context, but inactive when encoding standalone corpus items.

For SFT, the pretrained encoder retrieves a top-10 feedback pool for each query. We sample three items without replacement to construct four query states. Missing positives are not added to this pool, and relevance labels are not used to filter feedback. The annotated positive is included separately in the contrastive supervision set.
We train the SFT model for two epochs. In the first epoch, we disable the degradation penalty by setting $\lambda_{\text{deg}}=0$ in Eq.~\ref{eq:sft-objective}, allowing the model to first acquire basic multi-round retrieval capability. In the second epoch, we enable the degradation penalty to encourage the model to leverage retrieved items to refine the query representation.

For RL, we sample 2K queries from each of the 16 M-BEIR dataset--task combinations, giving 32K queries. We form $\mathcal{C}_{\mathrm{RL}}(q)=\mathcal{C}_{100}(q)\cup\{d^+\}$ from the SFT model's top-100 results and the annotated positive, removing duplicates. The resulting pool is shared by feedback sampling, reward computation, and final-step contrastive supervision. The history mask applies only to feedback selection. The reward gate in \eqref{eq:final-supervision} is applied to each trajectory before averaging its auxiliary loss. Unless otherwise stated, $N=4$ at both training and inference, and inference selects the highest-scoring unvisited feedback item at each step.

For queries with multiple positive items, during SFT we randomly select one positive as the ground-truth target and mask the remaining positives. During RL, the final-embedding supervision averages the losses over all positive items, while the reward function applies max pooling over their rewards.

% \paragraph{Optimization.}
Tables~\ref{tab:app-sft-settings} and~\ref{tab:app-rl-settings} list settings shared by ARR-2B and ARR-8B. SFT batch sizes count query--positive pairs. RL rollout batch sizes count queries, whereas optimization batch sizes count trajectories.
% Policy sampling uses a temperature of $\tau=0.1$ in Eq.~\ref{}; the contrastive temperatures are $\tau=0.02$ in SFT and $\tau=0.05$ for final-step supervision in RL. We use the final checkpoint from each training stage.

% \paragraph{Computational resources.}
SFT takes 24 hours on 8 NVIDIA H20 GPUs for the 2B model and 48 hours on 16 NVIDIA H20 GPUs for the 8B model. RL uses 8 NVIDIA H20 GPUs for both models, taking 8 hours for 2B and 24 hours for 8B.

\begin{table*}[t]
\centering
\footnotesize
\setlength{\tabcolsep}{3pt}
\renewcommand{\arraystretch}{1.16}
% Source: U-MARVEL, Appendix ``M-BEIR Dataset & Metrics'', tab:mbeir_overview.
% Do not import its aggregate training count: the prose and table disagree.
\caption{M-BEIR tasks and rounded benchmark statistics~\citep{M-BEIR,U-marvel}. Here, t denotes text and i denotes image; i+t denotes a joint image--text input. Split sizes count queries; ``Pool'' counts candidates in the local evaluation collection. These are benchmark statistics, not counts after ARR-specific preprocessing.}
\label{tab:app-mbeir-data}
\begin{tabular}{lccc}
\toprule
Dataset & Query $\to$ item & Train / dev / test queries & Pool \\
\midrule
VisualNews (VN)~\citep{liu2020visual} & t $\to$ i  & 99K / 20K / 20K & 542K \\
MSCOCO (CO)~\citep{lin2014microsoft} & t $\to$ i  & 100K / 24.8K / 24.8K & 5K \\
Fashion200K (F200)~\citep{han2017automatic} & t $\to$ i & 15K / 1.7K / 1.7K & 201K \\
WebQA (WQ)~\citep{chang2022webqa} & t $\to$ t  & 16K / 1.7K / 2.4K & 544K \\
EDIS (ES)~\citep{liu2023edis} & t $\to$ i+t &  26K / 3.2K / 3.2K & 1M \\
WebQA (WQ) & t $\to$ i+t & 17K / 1.7K / 2.5K & 403K \\
VisualNews (VN) & i $\to$ t & 100K / 20K / 20K & 537K \\
MSCOCO (CO) & i $\to$ t  & 113K / 5K / 5K & 25K \\
Fashion200K (F200) & i $\to$ t & 15K / 4.8K / 4.8K & 61K \\
NIGHTS (NS)~\citep{fu2023dreamsim} & i $\to$ i  & 16K / 2K / 2K & 40K \\
OVEN (ON)~\citep{hu2023open} & i+t $\to$ t  & 150K / 50K / 50K & 676K \\
InfoSeek (InS)~\citep{chen2023can} & i+t $\to$ t  & 141K / 11K / 11K & 611K \\
FashionIQ (FIQ)~\citep{wu2021fashion} & i+t $\to$ i  & 16K / 2K / 6K & 74K \\
CIRR (CR)~\citep{liu2021image} & i+t $\to$ i &  26K / 2K / 4K & 21K \\
OVEN (ON) & i+t $\to$ i+t & 157K / 14.7K / 14.7K & 335K \\
InfoSeek (InS) & i+t $\to$ i+t & 143K / 17.6K / 17.6K & 481K \\
\bottomrule
\end{tabular}
\end{table*}

\begin{table*}[t]
\centering
\footnotesize
\setlength{\tabcolsep}{1pt}
\renewcommand{\arraystretch}{1.12}
% Source: U-MARVEL, tab:zero_shot_benchmark_summary and its main zero-shot table.
% Its Flickr30K 1K-query/5K-candidate entry describes image-to-text retrieval;
% reverse the counts for text-to-image retrieval over the same image-caption set.
\caption{Zero-shot evaluation tasks and metrics. Counts are listed separately when retrieval directions differ. Candidate counts refer to the pool searched by each query; GeneCIS uses query-specific candidate sets.}
\label{tab:app-zero-data}
\begin{tabular}{lcc}
\toprule
Benchmark & Query $\to$ item & Queries / candidates  \\
\midrule
ShareGPT4V (Share4V)~\citep{chen2024sharegpt4v} & t $\to$ i; i $\to$ t & 1K / 1K (both directions)\\
Urban-1K (Urban)~\citep{zhang2024long} & t $\to$ i; i $\to$ t & 1K / 1K (both directions)  \\
Flickr30K (Flickr)~\citep{plummer2015flickr30k} & t $\to$ i; i $\to$ t & t $\to$ i: 5K / 1K; i $\to$ t: 1K / 5K  \\
CIRCO~\citep{baldrati2023zero} & i+t $\to$ i & 800 / 120K \\
GeneCIS~\citep{vaze2023genecis} & i+t $\to$ i & 8K / 10--15 \\
Visual Dialog (VisD)~\citep{das2017visual} & dialogue $\to$ i & 2K / 2K  \\
Multi-round FashionIQ (MT-FIQ)~\citep{yuan2021conversational} & interactive i/t $\to$ i & 2.4K / 6.2K  \\
\bottomrule
\end{tabular}
\end{table*}

\begin{table}[!htbp]
\centering
\footnotesize
\setlength{\tabcolsep}{6pt}
\renewcommand{\arraystretch}{1.08}
\caption{SFT training settings shared by ARR-2B and ARR-8B.}
\label{tab:app-sft-settings}
\begin{tabular*}{0.9\linewidth}{@{\extracolsep{\fill}}lc@{}}
\toprule
\textbf{Setting} & \textbf{Value} \\
\midrule
% \multicolumn{2}{@{}l}{\textit{Optimization}} \\
Optimizer & AdamW \\
Weight decay & 0.01 \\
Learning rate & $5\times10^{-5}$ \\
Learning-rate schedule & Cosine \\
Warmup steps & 150 \\
Training epochs & 2 \\
% Gradient accumulation steps & 1 \\
Effective global batch size (pairs) & 1,024 \\
Gradient clipping (maximum norm) & 1.0 \\
\midrule
% \multicolumn{2}{@{}l}{\textit{Loss}} \\
Contrastive temperature ($\tau$ in Eq.~\ref{eq:infonce})& 0.02 \\
Degradation penalty weight $\lambda_{\mathrm{deg}}$ & 0.1 \\
Hard negatives & 0 \\
\midrule
% \multicolumn{2}{@{}l}{\textit{Input and execution}} \\
Maximum text tokens & 1,024 \\
Maximum image pixels & $1{,}280\times32\times32$ \\
Minimum image pixels & $64\times32\times32$ \\
Training precision & BF16 \\
Checkpoint selection & Final \\
\bottomrule
\end{tabular*}
\end{table}

\begin{table}[!htbp]
\centering
\footnotesize
\setlength{\tabcolsep}{6pt}
\renewcommand{\arraystretch}{1.08}
\caption{RL training settings shared by ARR-2B and ARR-8B. All other settings follow those used for SFT.}
\label{tab:app-rl-settings}
\begin{tabular*}{0.9\linewidth}{@{\extracolsep{\fill}}lc@{}}
\toprule
\textbf{Setting} & \textbf{Value} \\
\midrule
% \multicolumn{2}{@{}l}{\textit{Optimization}} \\
Learning rate & $5\times10^{-6}$ \\
Learning-rate schedule & Cosine \\
Warmup steps & 50 \\
Training epochs & 2 \\
Effective global batch size (trajectories) & 256 \\
\midrule
% \multicolumn{2}{@{}l}{\textit{LoRA configuration}} \\
LoRA rank & 128 \\
LoRA scaling $\alpha$ & 256 \\
LoRA dropout & 0.0 \\
\midrule
% \multicolumn{2}{@{}l}{\textit{Rollouts and loss}} \\
Global rollout batch size (queries) & 1,024 \\
Trajectories per query $M$ & 8 \\
Policy sampling temperature ($\tau$ in Eq.~\ref{eq:feedback-policy})  & 0.1 \\
GRPO clipping $\epsilon$ & 0.2 \\
Optimization passes per rollout batch & 1 \\
Final-step contrastive temperature ($\tau$ in Eq.~\ref{eq:final-supervision})  & 0.05 \\
Final-step contrastive weight $\lambda_{\mathrm{ret}}$ & 0.1 \\
Reward threshold $R$ & 0.2 \\
\bottomrule
\end{tabular*}
\end{table}

\subsection{Baselines}

\paragraph{MM-Embed.}
MM-Embed~\citep{MM-Embed} adapts an MLLM for universal multimodal retrieval through contrastive learning and modality-aware hard-negative mining, with text retrieval data providing additional supervision.

\paragraph{LamRA-Ret.}
LamRA~\citep{Lamra} provides separate retrieval and reranking components. Its retrieval component, LamRA-Ret, learns multimodal embeddings through language-only pretraining and multimodal instruction tuning. We compare with the retrieval component in our main tables.

\paragraph{TRACE.}
TRACE~\citep{TRACE} integrates reasoning and representation learning in an MLLM, adapting its reasoning behavior to the retrieval task. It represents an approach that improves query interpretation before embedding.

\paragraph{ELVA.}
ELVA~\citep{ELVA} constructs embeddings after generating a textual synthesis of the input and uses ranking and margin rewards for RL. Its policy training targets retrieval-oriented representations, providing a comparison with ARR's policy over feedback items.

\section{Hyperparameter Analysis}
\label{sec:hyperparameters}
\begin{table}[!htbp]
\centering
\caption{
Sensitivity to auxiliary loss weights for ARR-2B on M-BEIR (average recall, \%).
Bold marks the best result in each column within each training stage.
}
\label{tab:hyper}
\small
\setlength{\tabcolsep}{4pt}
\renewcommand{\arraystretch}{1.12}

\begin{tabular*}{0.9\linewidth}{@{\extracolsep{\fill}}l*{4}{c}@{}}
\toprule
\textbf{Setting} & \textbf{Iter-1} & \textbf{Iter-2} & \textbf{Iter-3} & \textbf{Iter-4} \\
\midrule
\multicolumn{5}{@{}l}{\textit{ARR-SFT}} \\
$\lambda_\text{deg}=0$ & \textbf{55.69} & 55.61 & 55.66 & 55.68 \\
$\lambda_\text{deg}=0.1$ (default) & 55.63 & \textbf{55.95} & \textbf{56.01} & \textbf{56.01} \\
$\lambda_\text{deg}=1.0$ & 55.60 & 55.76 & 55.86 & 55.89 \\
\midrule
\multicolumn{5}{@{}l}{\textit{ARR-RL}} \\
$\lambda_\text{ret}=0$ & 56.20 & 56.40 & 56.39 & 56.43 \\
$\lambda_\text{ret}=0.01$ & 56.15 & 56.45 & 56.46 & 56.47 \\
$\lambda_\text{ret}=0.1$ (default) & \textbf{56.36} & \textbf{56.68} & \textbf{56.66} & \textbf{56.74} \\
$\lambda_\text{ret}=1.0$ & 55.59 & 55.92 & 56.01 & 56.02 \\
\bottomrule
\end{tabular*}
\end{table}

Table~\ref{tab:hyper} examines the degradation-penalty weight $\lambda_{\mathrm{deg}}$ in \eqref{eq:sft-objective} and the final-step supervision weight $\lambda_{\mathrm{ret}}$ in \eqref{eq:rl-objective}, using ARR-2B. We use 0.1 for both weights in the default configuration.

\paragraph{Degradation penalty.}
Without the penalty ($\lambda_{\mathrm{deg}}=0$), the score remains nearly constant across inference states: 55.69 at Iter-1 and 55.68 at Iter-4. With $\lambda_{\mathrm{deg}}=0.1$, it rises from 55.63 to 56.01. Increasing the weight to 1.0 also produces improvements across states, but yields a lower final score of 55.89. Thus, greater emphasis on preventing relative degradation does not necessarily maximize absolute retrieval accuracy. Among the tested values, 0.1 gives the best final score while maintaining comparable initial performance.

\paragraph{Final-step contrastive supervision.}
Without the auxiliary loss ($\lambda_{\mathrm{ret}}=0$), policy training reaches 56.43 at Iter-4, above the SFT score of 56.01. Setting $\lambda_{\mathrm{ret}}=0.1$ further raises this score to 56.74. A weight of 1.0 yields 56.02, suggesting that excessive emphasis on the auxiliary objective can reduce the gains from policy training. The intermediate value 0.1 performs best among the tested settings.

\end{document}